\documentclass[a4paper]{article}

\usepackage{INTERSPEECH2019}
\usepackage{multirow}
\usepackage{url}
\usepackage{color}
\usepackage{todonotes}
\usepackage{verbatim}
\usepackage{verbatim}

\usepackage{caption} 
\usepackage{xcolor}
\usepackage{colortbl}
\definecolor{DNN}{RGB}{220,220,220}
\definecolor{DNNColor}{RGB}{220,220,220}
\definecolor{FusionColor}{RGB}{181,167,151}

\newcommand{\reducedstrut}{\vrule width 0pt height .9\ht\strutbox depth .9\dp\strutbox\relax}
\newcommand{\dnncolor}[1]{%
  \begingroup
  \setlength{\fboxsep}{0pt}%  
  \colorbox{DNNColor}{\reducedstrut#1\/}%
  \endgroup
}

\title{ASVspoof 2019: Future Horizons in Spoofed and Fake Audio Detection}

\name{Massimiliano Todisco$^1$, Xin Wang$^2$, Ville Vestman$^{3,6}$, Md Sahidullah$^4$, H\'ector Delgado$^1$,\\ Andreas Nautsch$^1$, Junichi Yamagishi$^{2,5}$, Nicholas Evans$^1$, Tomi Kinnunen$^3$ and Kong Aik Lee$^6$}
\address{
  $^1$EURECOM, France -- $^2$National Institute of Informatics, Japan -- $^3$University of Eastern Finland, Finland -- $^4$Inria, France -- $^5$University of Edinburgh, UK -- $^6$NEC Corp., Japan}
\email{info@asvspoof.org}
\ninept

\begin{document}
\maketitle
\begin{abstract}
ASVspoof, now in its third edition, is a series of community-led challenges which promote the development of countermeasures to protect automatic speaker verification (ASV) from the threat of spoofing.  Advances in the 2019 edition include: (i) a consideration of both logical access (LA) and physical access (PA) scenarios and the three major forms of spoofing attack, namely synthetic, converted and replayed speech; (ii) spoofing attacks generated with state-of-the-art neural acoustic and waveform models; (iii) an improved, controlled simulation of replay attacks; (iv) use of the tandem detection cost function (t-DCF) that reflects the impact of both spoofing and countermeasures upon ASV reliability.  Even if ASV remains the core focus, in retaining the equal error rate (EER) as a secondary metric, ASVspoof also embraces the growing importance of {\it fake audio detection}.  ASVspoof 2019 attracted the participation of 63 research teams, with more than half of these reporting systems that improve upon the performance of two baseline spoofing countermeasures. This paper describes the 2019 database, protocols and challenge results.  It also outlines major findings which demonstrate the real progress made in protecting against the threat of spoofing and fake audio.

\end{abstract}
\noindent\textbf{Index Terms}: spoofing, automatic speaker verification, ASVspoof, presentation attack detection, fake audio.

\section{Introduction}
\vspace{-1mm}

The ASVspoof initiative~\cite{interspeechSpecialSession2013, Wu-ASVspoof2015,Kinnunen2017-assessing} spearheads research in anti-spoofing for automatic speaker verification (ASV). Previous ASVspoof editions focused on the design of spoofing countermeasures for synthetic and converted speech (2015) and replayed speech (2017). ASVspoof 2019~\cite{ASVspoof19_evalplan}, the first edition to focus on all three major spoofing attack types, extends previous challenges in several directions, not least in terms of adressing two different use case scenarios: logical access (LA) and physical access (PA).  

The LA scenario involves spoofing attacks that are injected directly into the ASV system.  Attacks in the LA scenario are generated using the latest text-to-speech synthesis (TTS) and voice conversion (VC) technologies. The best of these algorithms produce speech which is perceptually indistinguishable from bona fide speech.  ASVspoof 2019 thus aims to determine whether the advances in TTS and VC technology pose a greater threat to the reliability of ASV systems, as well as spoofing countermeasures.
For the PA scenario, speech data is assumed to be captured by a microphone in a physical, reverberant space.  Replay spoofing attacks are recordings of bona fide speech which are assumed to be captured, possibly surreptitiously, and then re-presented to the microphone of an ASV system using a replay device.  In contrast to the 2017 edition of ASVspoof, the 2019 edition PA database is constructed from a far more controlled simulation of replay spoofing attacks that is also relevant to the study of fake audio detection in the case of, \emph{e.g.}\ smart home devices.

While the \emph{equal error rate} (EER) metric of previous editions is retained as a secondary metric, ASVspoof 2019 migrates to a new primary metric in the form of the ASV-centric tandem decision cost function (t-DCF)~\cite{Kinnunen2018-tDCF}. While the challenge is still a stand-alone spoofing detection task which does not require expertise in ASV, adoption of the t-DCF ensures that scoring and ranking reflects the comparative impact of both spoofing and countermeasures \emph{upon an ASV system}.

This paper describes the ASVspoof 2019 challenge, the LA and PA scenarios, the evaluation rules and protocols, the t-DCF metric, the common ASV system, baseline countermeasures and challenge results.

\section{Database}
\vspace{-1mm}

The ASVspoof 2019 database encompasses two partitions for the assessment of LA and PA scenarios.  Both are derived from the VCTK base corpus\footnote{http://dx.doi.org/10.7488/ds/1994} which includes speech data captured from 107 speakers (46 males, 61 females). Both LA and PA databases are themselves partitioned into three datasets, namely training, development and evaluation which comprise the speech from 20 (8 male, 12 female), 10 (4 male, 6 female) and 48 (21 male, 27 female) speakers respectively.  The three partitions are disjoint in terms of speakers and the recording conditions for all source data are identical.  While the training and development sets contain spoofing attacks generated with the same algorithms/conditions (designated as \emph{known} attacks), the evaluation set also contains attacks generated with different algorithms/conditions (designated as \emph{unknown} attacks).  Reliable spoofing detection performance therefore calls for systems that generalise well to previously-unseen spoofing attacks. With full descriptions available in the ASVspoof 2019 evaluation plan~\cite{ASVspoof19_evalplan}, the following presents a summary of the specific characteristics of the LA and PA databases.

\vspace{-1mm}
\subsection{Logical access}
\vspace{-1mm}

The LA database contains bona fide speech and spoofed speech data generated using 17 different TTS and VC systems. Data used for the training of TTS and VC systems also comes from the VCTK database but there is no overlap with the data contained in the 2019 database. Six of these systems are designated as known attacks, with the other 11 being designated as unknown attacks. The training and development sets contain known attacks only whereas the evaluation set contains 2 known and 11 unknown spoofing attacks. Among the 6 known attacks there are 2 VC systems and 4 TTS systems.  VC systems use a neural-network-based and spectral-filtering-based approaches~\cite{matrouf2006effect}.  TTS systems use either waveform concatenation or neural-network-based speech synthesis using a conventional source-filter vocoder~\cite{morise2016world} or a WaveNet-based vocoder~\cite{oord2016wavenet}. The 11 unknown systems comprise 2 VC, 6 TTS and 3 hybrid TTS-VC systems and were implemented with various waveform generation methods including classical vocoding, Griffin-Lim~\cite{griffin1984signal}, generative adversarial networks~\cite{tanaka2018synthetic}, neural waveform models~\cite{oord2016wavenet,wang2018neural}, waveform concatenation, waveform filtering~\cite{kobayashi2014statistical}, spectral filtering, and their combination.

\vspace{-1mm}
\subsection{Physical access}
\vspace{-1mm}

Inspired by work to analyse and improve ASV reliability in reverberant conditions~\cite{Ko2017ASO,Snyder2018XVectorsRD} and a similar approach used in the study of replay reported in~\cite{janicki2016assessment}, both bona fide data and spoofed data contained in the PA database are generated according to a simulation~\cite{campbell05,vincent08,novak2015synchronized} of their presentation to the microphone of an ASV system within a reverberant acoustic environment.  Replayed speech is assumed first to be captured with a recording device before being replayed using a non-linear replay device. Training and development data is created according to 27 different acoustic and 9 different replay configurations. Acoustic configurations comprise an exhaustive combination of 3 categories of room sizes, 3 categories of reverberation and 3 categories of speaker/talker\footnote{From hereon we refer to \textit{talkers} in order to avoid potential confusion with loud\emph{speakers} used to mount replay spoofing
attacks.}-to-ASV microphone distances. Replay configurations comprise 3 categories of attacker-to-talker recording distances, and 3 categories of loudspeaker quality.  Replay attacks are simulated with a random attacker-to-talker recording distance and a random loudspeaker quality corresponding to the given configuration category.  Both bona fide and replay spoofing access attempts are made with a random room size, reverberation level and talker-to-ASV microphone distance.

Evaluation data is generated in the same manner as training and development data, albeit with different, random acoustic and replay configurations.  The set of room sizes, levels of reverberation, talker-to-ASV microphone distances, attacker-to-talker recording distances and loudspeaker qualities, while drawn from the same configuration categories, are different to those for the training and development set.   Accordingly, while the categories are the same and \emph{known} a priori, the specific impulse responses and replay devices used to simulate bona fide and replay spoofing access attempts are different or \emph{unknown}.  It is expected that reliable performance will only be obtained by countermeasures that generalise well to these conditions, i.e.\ countermeasures that are not over-fitted to the specific acoustic and replay configurations observed in training and development data.

\section{Performance measures and baselines}
\vspace{-1mm}

ASVspoof 2019 focuses on assessment of \emph{tandem} systems consisting of both a spoofing countermeasure (CM) (designed by the participant) and an ASV system (provided by the organisers). The performance of the two combined systems is evaluated via the minimum normalized \textbf{tandem detection cost function} (t-DCF, for the sake of easier tractability)~\cite{Kinnunen2018-tDCF} of the form:
\begin{equation}
\text{t-DCF}_\text{norm}^\text{min}=\min_s \left\{\,\beta P_\text{miss}^\text{cm}(s)+P_\text{fa}^\text{cm}(s)\,\right\},
\label{equ:tdcf}
\end{equation}
\noindent where $\beta$ depends on application parameters (priors, costs) \emph{and} ASV performance (miss, false alarm, and spoof miss rates), while $P_\text{miss}^\text{cm}(s)$ and $P_\text{fa}^\text{cm}(s)$ are the CM miss and false alarm rates at threshold $s$. The minimum in \eqref{equ:tdcf} is taken over all thresholds on given data (development or evaluation) with a known key, corresponding to oracle calibration. While the challenge rankings are based on \emph{pooled} performance in either scenario (LA or PA), results are also presented when decomposed by attack. In this case, $\beta$ depends on the effectiveness of each attack. In particular, with everything else being constant, $\beta$ is \emph{inversely proportional to the ASV false accept rate for a specific attack}: the penalty when a CM falsely rejects bona fide speech is higher in the case of less effective attacks. Likewise, the relative penalty when a CM falsely accepts spoofs is higher for more effective attacks. Thus, while \eqref{equ:tdcf} appears to be deceptively similar to the NIST DCF, $\beta$ (hence, the cost function itself) is automatically adjusted according to the effectiveness of each attack. 
Full details of the t-DCF metric and specific configuration parameters as concerns ASVspoof 2019 are presented in~\cite{ASVspoof19_evalplan}. The EER serves as a secondary metric.  The EER corresponds to a CM operating point with equal miss and false alarm rates and was the primary metric for previous editions of ASVspoof.  Without an explicit link to the impact of CMs upon the reliability of an ASV system, the EER may be more appropriate as a metric for fake audio detection, \emph{i.e.}\ where there is no ASV system.

The common ASV system uses \emph{x-vector} speaker embeddings~\cite{Snyder2018XVectorsRD} together with a \emph{probabilistic linear discriminant analysis} (PLDA)~\cite{prince2007probabilistic} backend. The x-vector model used to compute ASV scores required to compute the t-DCF  is pre-trained\footnote{\url{http://kaldi-asr.org/models/m7}} with the Kaldi~\cite{povey2011kaldi} VoxCeleb~\cite{nagrani2017voxceleb} recipe. The original recipe is modified to include PLDA adaptation using disjoint, bona fide, in-domain data. Adaptation was performed separately for LA and PA scenarios since bona fide recordings for the latter contain additional simulated acoustic and recording effects. The ASV operating point, needed in computing $\beta$ in \eqref{equ:tdcf}, is set to the EER point based on target and nontarget scores.

ASVspoof 2019 adopted two CM baseline systems. They use a common Gaussian mixture model (GMM) back-end classifier with either \emph{constant Q cepstral coefficient} (CQCC) features~\cite{CQCC2016,TODISCO2017516} (B01) or \emph{linear frequency cepstral coefficient} (LFCC) features~\cite{Sahidullah15} (B02). 

\section{Challenge results}
\vspace{-1mm}
  
\begin{table}[!t]
\caption{Primary system results. Results shown in terms of minimum t-DCF and the CM EER [\%]. 
IDs highlighted in \dnncolor{grey} signify systems that used neural networks in either the front- or back-end. IDs highlighted in \textbf{bold font} signify systems that use an ensemble of classifiers.}
\scriptsize
\tabcolsep 5.5pt
\vspace{-2mm}
\begin{center}
\begin{tabular}{|p{.1cm}ccp{.5cm}|p{.1cm}ccp{.6cm}|}
\hline
\multicolumn{8}{|c|}{\footnotesize{\textbf{ASVspoof 2019 LA scenario}}} \\ \hline
\textbf{\#} & \textbf{ID} & \textbf{t-DCF} & \textbf{EER} & \textbf{\#} & \textbf{ID} & \textbf{t-DCF} & \textbf{EER} \\ \hline
1 & \cellcolor{DNN} \textbf{T05} & 0.0069 & 0.22 & 	26 & \cellcolor{DNN} T57 & 0.2059 & 10.65 \\ \hline
2 & \cellcolor{DNN} \textbf{T45} & 0.0510 & 1.86 & 	27 & \cellcolor{DNN} \textbf{T42} & 0.2080 & 8.01 \\ \hline
3 & \cellcolor{DNN} \textbf{T60} & 0.0755 & 2.64 & 	28 & \textit{B02} & {0.2116} & {8.09} \\ \hline
4 & \cellcolor{DNN} \textbf{T24} & 0.0953 & 3.45 & 	29 & \textbf{T17} & 0.2129 & 7.63 \\ \hline
5 & \cellcolor{DNN} \textbf{T50} & 0.1118 & 3.56 & 	30 & \cellcolor{DNN} \textbf{T23} & 0.2180 & 8.27 \\ \hline
6 & \cellcolor{DNN} \textbf{T41} & 0.1131 & 4.50 & 	31 & \cellcolor{DNN} \textbf{T53} & 0.2252 & 8.20 \\ \hline
7 & \cellcolor{DNN} \textbf{T39} & 0.1203 & 7.42 & 	32 & \textbf{T59} & 0.2298 & 7.95 \\ \hline
8 & \textbf{T32} & 0.1239 & 4.92 & 				33 & \textit{B01} & {0.2366} & {9.57} \\ \hline
9 & \textbf{T58} & 0.1333 & 6.14 & 				34 & T52 & 0.2366 & 9.25 \\ \hline
10 & T04 & 0.1404 & 5.74 & 					35 & \cellcolor{DNN} \textbf{T40} & 0.2417 & 8.82 \\ \hline
11 & \textbf{T01} & 0.1409 & 6.01 & 				36 & T55 & 0.2681 & 10.88 \\ \hline
12 & \cellcolor{DNN}\textbf{T22} & 0.1545 & 6.20 & 	37 & \textbf{T43} & 0.2720 & 13.35 \\ \hline
13 & \cellcolor{DNN}T02 & 0.1552 & 6.34 & 		38 & \cellcolor{DNN} T31 & 0.2788 & 15.11 \\ \hline
14 & \cellcolor{DNN}\textbf{T44} & 0.1554 & 6.70 & 	39 & \cellcolor{DNN} \textbf{T25} & 0.3025 & 23.21 \\ \hline
15 & \cellcolor{DNN}\textbf{T16} & 0.1569 & 6.02 & 	40 & \cellcolor{DNN} \textbf{T26} & 0.3036 & 15.09 \\ \hline
16 & T08 & 0.1583 & 6.38 & 					41 & \cellcolor{DNN} T47 & 0.3049 & 18.34 \\ \hline
17 & \cellcolor{DNN} \textbf{T62} & 0.1628 & 6.74 & 	42 & \cellcolor{DNN} T46 & 0.3214 & 12.59 \\ \hline
18 &  \textbf{T27} & 0.1648 & 6.84 & 			43 & T21 & 0.3393 & 19.01 \\ \hline
19 &  \textbf{T29} & 0.1677 & 6.76 & 			44 & T61 & 0.3437 & 15.66 \\ \hline
20 & \cellcolor{DNN} \textbf{T13} & 0.1778 & 6.57 & 	45 & \cellcolor{DNN} \textbf{T11} & 0.3742 & 18.15 \\ \hline
21 & \cellcolor{DNN} \textbf{T48} & 0.1791 & 9.08 & 	46 & \cellcolor{DNN} \textbf{T56} & 0.3856 & 15.32 \\ \hline
22 & \cellcolor{DNN} \textbf{T10} & 0.1829 & 6.81 & 	47 & T12 & 0.4088 & 18.27 \\ \hline
23 & \cellcolor{DNN} T54 & 0.1852 & 7.71 & 		48 & \cellcolor{DNN} T14 & 0.4143 & 20.60 \\ \hline
24 & {T38} & 0.1940 & 7.51 & 	49 & \cellcolor{DNN} T20 & 1.0000 & 92.36 \\ \hline
25 & \cellcolor{DNN} T33 & 0.1960 & 8.93 & 		50 & \cellcolor{DNN} T30 & 1.0000 & 49.60 \\ \hline
\end{tabular} \\
\end{center}

\vspace{-5mm}
\begin{center}
\begin{tabular}{|p{.1cm}ccp{.5cm}|p{.1cm}ccp{.6cm}|}
\hline
\multicolumn{8}{|c|}{\textbf{\footnotesize{ASVspoof 2019 PA scenario}}} \\ \hline
\textbf{\#} & \textbf{ID} & \textbf{t-DCF} & \textbf{EER} & \textbf{\#} & \textbf{ID} & \textbf{t-DCF} & \textbf{EER} \\ \hline
1 & \cellcolor{DNN} \textbf{T28} & 0.0096 & 0.39 & 	27 & \textbf{T29} & 0.2129 & 8.48 \\ \hline
2 & \cellcolor{DNN} \textbf{T45} & 0.0122 & 0.54 & 	28 & \textbf{T01} & 0.2129 & 9.07 \\ \hline
3 & \cellcolor{DNN} \textbf{T44} & 0.0161 & 0.59 & 	29 & \cellcolor{DNN} T54 & 0.2130 & 11.93 \\ \hline
4 & \cellcolor{DNN} \textbf{T10} & 0.0168 & 0.66 & 	30 & \cellcolor{DNN} T35 & 0.2286 & 7.77 \\ \hline
5 & \cellcolor{DNN} \textbf{T24} & 0.0215 & 0.77 & 	31 & \cellcolor{DNN} T46 & 0.2372 & 8.82 \\ \hline
6 & \cellcolor{DNN} \textbf{T53} & 0.0219 & 0.88 & 				32 & \textbf{T34} & 0.2402 & 10.35 \\ \hline
7 & \textbf{T17} & 0.0266 & 0.96 & 				33 & \textit{B01} & {0.2454} & {11.04} \\ \hline
8 & \cellcolor{DNN} \textbf{T50} & 0.0350 & 1.16 & 	34 & \cellcolor{DNN} \textbf{T38} & 0.2460 & 9.12 \\ \hline
9 & \cellcolor{DNN} \textbf{T42} & 0.0372 & 1.51 & 	35 & \cellcolor{DNN} \textbf{T59} & 0.2490 & 10.53 \\ \hline
10 & \cellcolor{DNN} \textbf{T07} & 0.0570 & 2.45 & 	36 & \cellcolor{DNN} T03 & 0.2593 & 11.26 \\ \hline
11 & \cellcolor{DNN} T02 & 0.0614 & 2.23 & 					37 & \cellcolor{DNN} \textbf{T51} & 0.2617 & 11.92 \\ \hline
12 & \cellcolor{DNN} \textbf{T05} & 0.0672 & 2.66 & 	38 & T08 & 0.2635 & 10.97 \\ \hline
13 & \cellcolor{DNN} \textbf{T25} & 0.0749 & 3.01 & 	39 & \textbf{T58} & 0.2767 & 11.28 \\ \hline
14 & \cellcolor{DNN} \textbf{T48} & 0.1133 & 4.48 & 	40 & \cellcolor{DNN} T47 & 0.2785 & 10.60 \\ \hline
15 & \cellcolor{DNN} T57 & 0.1297 & 4.57 & 	41 & \textbf{T09} & 0.2793 & 12.09 \\ \hline
16 & \cellcolor{DNN} \textbf{T31} & 0.1299 & 5.20 & 	42 & {T32} & 0.2810 & 12.20 \\ \hline
17 & \cellcolor{DNN} \textbf{T56} & 0.1309 & 4.87 & 	43 & T61 & 0.2958 & 12.53 \\ \hline
18 & \cellcolor{DNN} T49 & 0.1351 & 5.74 & 	44 & \textit{B02} & {0.3017} & {13.54} \\ \hline
19 & \cellcolor{DNN} \textbf{T40} & 0.1381 & 5.95 & 	45 & \cellcolor{DNN} \textbf{T62} & 0.3641 & 13.85 \\ \hline
20 & \cellcolor{DNN} \textbf{T60} & 0.1492 & 6.11 & 	46 & \cellcolor{DNN} T19 & 0.4269 & 21.25 \\ \hline
21 & \cellcolor{DNN} T14 & 0.1712 & 6.50 & 	47 & T36 & 0.4537 & 18.99 \\ \hline
22 & \cellcolor{DNN} \textbf{T23} & 0.1728 & 7.19 & 	48 & \cellcolor{DNN} \textbf{T41} & 0.5452 & 28.98 \\ \hline
23 & \cellcolor{DNN} \textbf{T13} & 0.1765 & 7.61 & 	49 & T21 & 0.6368 & 27.50 \\ \hline
24 & \textbf{T27} & 0.1819 & 7.98 & 				50 & T15 & 0.9948 & 42.28 \\ \hline
25 & \cellcolor{DNN} \textbf{T22} & 0.1859 & 7.44 & 	51 & \cellcolor{DNN} T30 & 0.9998 & 50.19 \\ \hline
26 & \textbf{T55} & 0.1979 & 8.19 & 				52 & \cellcolor{DNN} T20 & 1.0000 & 92.64 \\ \hline
\end{tabular}
\label{tab:participant-summary}
\end{center}
\vspace{-7mm}
\end{table}

Table~\ref{tab:participant-summary} shows results\footnote{As for previous editions of ASVspoof, results are anonymised, with individual teams being able to identify their position in the evaluation rankings via an identifier communicated separately to each of them.} in terms of the t-DCF and EER for primary systems, pooled over all attacks. For the LA scenario, 27 of the 48 participating teams produced systems that outperformed the best baseline B02.  For the PA scenario, the performance of B01 was bettered by 32 of the 50 participating teams. There is substantial variation in minimum t-DCF and EER for both LA and PA scenarios. The top-performing system for the LA scenario, T05, achieved a t-DCF of 0.0069 and EER of 0.22\%. The top-performing system for the PA scenario, T28, achieved a t-DCF of 0.0096 and EER of 0.39\%.  Confirming observations reported in~\cite{Kinnunen2018-tDCF}, monotonic increases in the t-DCF that are not always mirrored by monotonic increases in the EER show the importance of considering the performance of the ASV and CM systems \emph{in tandem}.
Table~\ref{tab:participant-summary} also shows that the top 7 (LA) and 6 (PA) systems used neural networks whereas 9 (LA) and 10 (PA) systems used an ensemble of classifiers.

\begin{figure}[!t]
    \vspace{-8mm}
	\centering
	\includegraphics[width=7cm,clip]{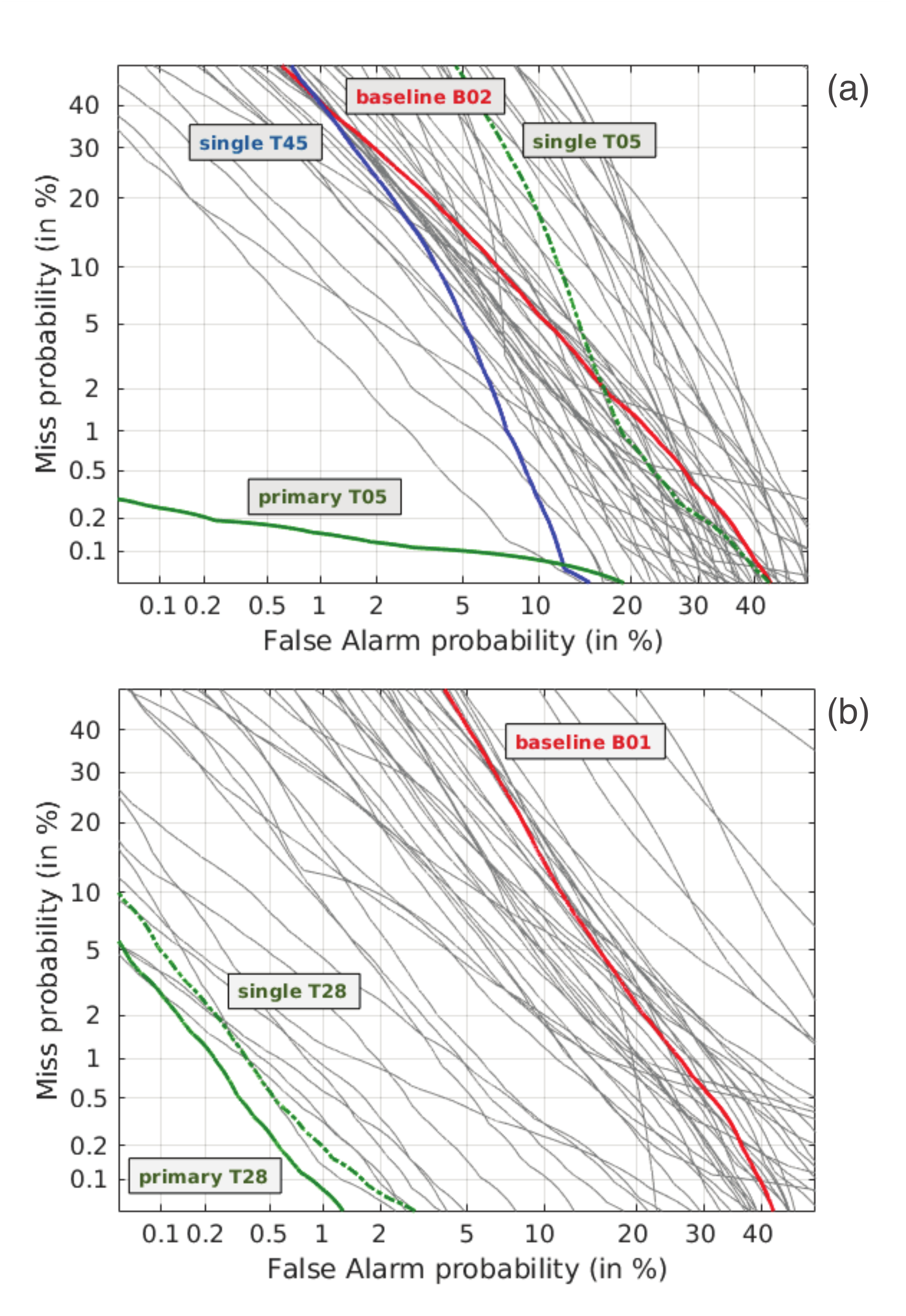}
	\vspace{-2mm}
	\caption{CM DET profiles for (a) LA and (b) PA scenarios.}
	\label{fig:DET}
	\vspace{-3mm}
\end{figure}

\vspace{-1mm}
\subsection{CM analysis}
\vspace{-1mm}

Corresponding CM detection error trade-off (DET) plots (no combination with ASV) are illustrated for LA and PA scenarios in Fig.~\ref{fig:DET}.
Highlighted in both plots are profiles for the two baseline systems B01 and B02, the best performing primary systems for teams T05 and T28, and the same teams' single systems.  Also shown are profiles for the overall best performing single system for the LA and PA scenarios submitted by teams T45 and, again, T28 respectively.  
For the LA scenario, very few systems deliver EERs below 5\%.  A dense concentration of systems deliver EERs between 5\% and 10\%.  Of interest is the especially low EER delivered by the primary T05 system, which delivers a substantial improvement over the same team's best performing single system.  Even the overall best performing single system of T45 is some way behind, suggesting that reliable performance for the LA scenario depends upon the fusion of complementary sub-systems.  This is likely due to the diversity in attack families, namely TTS, VC and hybrid TTS-VC systems. 
Observations are different for the PA scenario.  There is a greater spread in EERs and the difference between the best performing primary and single systems (both from T28) is much narrower.  That a low EER can be obtained with a single system suggests that reliable performance is less dependent upon effective fusion strategies.  This might be due to lesser variability (as compared to that for the LA) in replay spoofing attacks; there is only one family of attack which exhibits differences only in the level of convolutional channel noise.

\begin{figure*}[h]
% 	\centering
\hspace*{-1.2cm} 
	\includegraphics[width=19.0cm,clip]{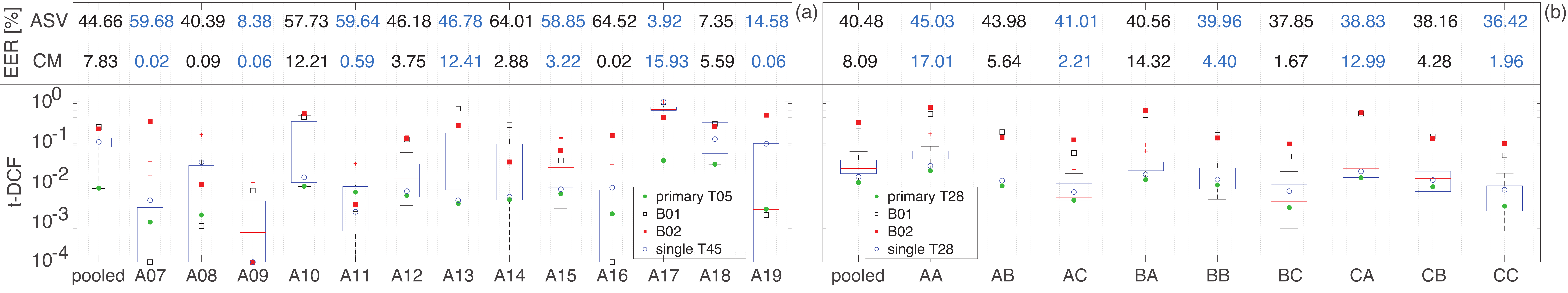}
	\vspace{-5mm}
	\caption{Boxplots of the top-10 performing LA (a) and PA (b) ASVspoof 2019 submissions. Results illustrated in terms of t-DCF decomposed for the 13 (LA) and 9 (PA) attacks in the evaluation partion. ASV under attack and the median CM EER [\%] of all the submitted systems are shown above the boxplots. A16 and A19 are known attacks.}
	\label{fig:BP}
\end{figure*}

\vspace{-1mm}
\subsection{Tandem analysis}
\vspace{-1mm}

Fig.~\ref{fig:BP} illustrates boxplots of the t-DCF when pooled (left-most) and when decomposed separately for each of the spoofing attacks in the evaluation set.  Results are shown individually for the best performing baseline, primary and single systems whereas the boxes illustrate the variation in performance for the top-10 performing systems.  Illustrated to the top of each boxplot are the EER of the common ASV system (when subjected to each attack) and the median CM EER across  all primary systems. The ASV system delivers baseline EERs (without spoofing attacks) of 2.48\% and 6.47\% for LA and PA scenarios respectively.

As shown in Fig.~\ref{fig:BP}(a) for the LA scenario, attacks A10, A13 and, to a lesser extent, A18, degrade ASV performance while being challenging to detect. They are end-to-end TTS with WaveRNN and a speaker encoder pretrained for ASV
\cite{DBLP:journals/corr/abs-1806-04558}, VC using moment matching networks \cite{li2015generative} and waveform filtering \cite{kobayashi2014statistical}, and i-vector/PLDA based VC \cite{kinnunen2017non} using a DNN glottal vocoder \cite{juvela2018speech}, respectively. 
Although A08, A12, and A15 also use neural waveform models and threaten ASV, they are easier to detect than A10. One reason may be that A08, A12, A15 are pipeline TTS and VC systems while A10 is optimized in an end-to-end manner. Another reason may be that A10 transfers ASV knowledge into TTS, implying that advances in ASV also improve the LA attacks. 
A17, a VAE-based VC \cite{Hsu2017} with waveform filtering, poses little threat to the ASV system, but it is the most difficult to detect and lead to the highest t-DCF. 
All the above attacks are new attacks not included in ASVspoof 2015. 

More consistent trends can be observed for the PA scenario. 
Fig.~\ref{fig:BP}(b) shows the t-DCF when pooled and decomposed for each of the 9 replay configurations.
Each attack is a combination of different attacker-to-talker recording distances~\{A,B,C\}X, and replay device qualities~X\{A,B,C\}~\cite{ASVspoof19_evalplan}. When subjected to replay attacks, the EER of the ASV system increases more when the attacker-to-talker distance is low (near-field effect) and when the attack is performed with higher quality replay devices (fewer channel effects). There are similar observations for CM performance and the t-DCF; lower quality replay attacks can be detected reliably whereas higher quality replay attacks present  more of a challenge.

\section{Discussion}
\vspace{-1mm}

Care must be exercised in order that t-DCF results are interpreted correctly. The reader may find it curious, for instance, that LA attack A17 corresponds to the \emph{highest} t-DCF while, with an ASV EER of 3.92\%, the attack is the \emph{least effective}. Conversely, attack A16 provokes an ASV EER of almost 65\%\footnote{Scores produced by spoofing attacks are higher than those of genuine trials.}, yet the median t-DCF is among the lowest. So, does A17 --- a weak attack --- really pose a problem? The answer is affirmative: A17 \emph{is} problematic, as far as the t-DCF is concerned. Further insight can be obtained from the attack-specific weights $\beta$ of \eqref{equ:tdcf}. For A17, a value of $\beta\approx 26$, indicates that the induced cost function provides 26 times higher penalty for rejecting bona fide users, than it does for missed spoofing attacks passed to the ASV system. The behavior of primary system T05 in Fig.~\ref{fig:DET}(a), with an aggressively tilted slope towards the low false alarm region, may explain why the t-DCF is near an order of magnitude better than the second best system.

\section{Conclusions}
\vspace{-1mm}

ASVspoof 2019 addressed two different spoofing scenarios, namely LA and PA, and also the three major forms of spoofing attack: synthetic, converted and replayed speech.  The LA scenario aimed to determine whether advances in countermeasure design have kept pace with progress in TTS and VC technologies and whether, as result, today's state-of-the-art systems pose a threat to the reliability of ASV. While findings show that the most recent techniques, e.g.\ those using neural waveform models and waveform filtering, in addition to those resulting from transfer learning (TTS and VC systems borrowing ASV techniques) do indeed provoke greater degradations in ASV performance, there is potential for their detection using countermeasures that combine multiple classifiers. The PA scenario aimed to assess the spoofing threat and countermeasure performance via simulation with which factors influencing replay spoofing attacks could be carefully controlled and studied.  Simulations consider variation in room size and reverberation time, replay device quality and the physical separation between both talkers and attackers (making surreptitious recordings) and talkers and the ASV system microphone.  Irrespective of the replay configuration, all replay attacks degrade ASV performance, yet, reassuringly, there is promising potential for their detection.

Also new to ASVspoof 2019 and with the objective of assessing the impact of both spoofing and countermeasures upon ASV reliability, is adoption of the ASV-centric t-DCF metric.  This strategy marks a departure from the independent assessment of countermeasure performance in isolation from ASV and a shift towards cost-based evaluation. Much of the spoofing attack research across different biometric modalities revolves around the premise that spoofing attacks are harmful and should be detected at any cost. That spoofing attacks have potential for harm is not in dispute.  It does not necessarily follow, however, that \emph{every} attack \emph{must} be detected. Depending on the application, spoofing attempts could be extremely rare or, in some cases, ineffective. Preparing for a worst case scenario, when that worst case is unlikely in practice, incurs costs of its own, \emph{i.e.}\ degraded user convenience.  The t-DCF framework enables one to encode explicitly the relevant statistical assumptions in terms of a well-defined cost function that generalises the classic NIST DCF. A key benefit is that the t-DCF disentangles the roles of ASV and CM developers as the error rates of the two systems are still treated independently.  As a result, ASVspoof 2019 followed the same, familiar format as previous editions, involving a low entry barrier --- participation still requires no ASV expertise and participants need submit countermeasures scores only --- the ASV system is provided by the organisers and is common to the assessment of all submissions.  With the highest number of submissions in ASVspoof's history, this strategy appears to have been a resounding success.  

\vspace{.2cm}
\noindent
\footnotesize
\textbf{Acknowledgements:}
The authors express their profound gratitude to the 27 persons from 14 organisations who contributed to creation of the LA database. The work was partially supported by: JST CREST Grant No.\ JPMJCR18A6 (VoicePersonae project), Japan; MEXT KAKENHI Grant Nos.\ (16H06302, 16K16096, 17H04687, 18H04120, 18H04112, 18KT0051), Japan; the VoicePersonae and RESPECT projects funded by the French Agence Nationale de la Recherche~(ANR); the Academy of Finland (NOTCH project\ no.\ 309629); Region Grand Est, France. The authors at the University of Eastern Finland also gratefully acknowledge the use of computational infrastructures at CSC -- IT Center for Science, and the support of NVIDIA Corporation with the donation of a Titan V GPU used in this research. 
\normalsize

\clearpage

\bibliographystyle{IEEEtran}
\bibliography{main}

\end{document}